\documentclass[a4paper]{jpconf}
\usepackage{graphicx}
\usepackage[hidelinks]{hyperref}
\usepackage{color}
\usepackage{amsmath,amsfonts,amssymb,bm}

\usepackage{lipsum}

\usepackage{fancyhdr}
\def\beq{\begin{equation}}
\def\eeq{\end{equation}}
\def\beqn{\begin{eqnarray}}
\def\eeqn{\end{eqnarray}}
\def\beqs{\begin{subequations}}
\def\eeqs{\end{subequations}}

\def\P {{\bf P}}

\def\bi {\mbox{\boldmath $i$}}

\def\P {{\bf P}}

\begin{document}

\title{Effects Beyond Center-of-Mass Separability in a Trapped Bosonic Mixture: Exact Results}

\author{O~E~Alon$^{1,2}$ and L~S~Cederbaum$^{3}$}
\address{$^{1}$ Department of Mathematics, University of Haifa, Haifa, Israel}
\address{$^{2}$ Haifa Research Center for Theoretical Physics and Astrophysics, University of Haifa, Haifa, Israel}
\address{$^{3}$ Theoretical Chemistry, Physical Chemistry Institute, Heidelberg University, Heidelberg, Germany}

\ead{ofir@research.haifa.ac.il}

\begin{abstract}
An exactly solvable model mimicking demixing
of two Bose-Einstein condensates
at the many-body level of theory is devised.
Various properties
are expressed in closed form along the demixing pathway and investigated. 
The connection between the center-of-mass coordinate and in particular the relative center-of-mass coordinate
and demixing is explained. 
The model is also exactly solvable at the mean-field level of theory,
allowing thereby comparison between
many-body and mean-field properties.
Applications are briefly discussed.
\end{abstract}

\section{Introduction}\label{INTRO}

Demixing of Bose-Einstein condensates has drawn an extensive attention and usually studied numerically,
either at the mean-field level or at the many-body level of theory, see, e.g.,
[1-29].
Spatial inhomogeneity makes the problem analytically almost intractable.
Here we introduce a solvable model which enables one to emulate
demixing, or mixing, of two Bose-Einstein condensates
at the many-body as well as at mean-field levels of theory.
With a solvable model we can investigate analytically various properties along the pathway of demixing,
such as the energetics, spatial overlap of the bosonic clouds,
and entanglement between the two species to list a few.
To this end, we extend the harmonic-interaction model for mixtures
[30-32]
to treat demixing.
Our work builds on and naturally goes beyond
[33-35].
The harmonic-interaction model for bosons, fermions, and mixtures has drawn much attention, see, e.g.,
[36-51].

\section{Theory and Properties}\label{THEORY}

Consider two Bose-Einstein condensates which consist of
species $1$ and species $2$ bosons, respectively.
Condensate $1$ is placed in an harmonic potential localized at the origin
and condensate $2$ is held in an harmonic potential centered at a distance $L$ from the first.
All particle-particle interactions are harmonic.
The many-particle Hamiltonian hence reads
\beqn\label{HAM_MIX_CAR}
& & \hat H(x_1,\ldots,x_{N_1},y_1,\ldots,y_{N_2}) = \nonumber \\
& & = \sum_{j=1}^{N_1} \left( -\frac{1}{2m_1} \frac{\partial^2}{\partial x_j^2} + \frac{1}{2} m_1\omega^2 x_j^2 \right)  
+ \sum_{j=1}^{N_2} \left[ -\frac{1}{2m_2} \frac{\partial^2}{\partial y_j^2} +
\frac{1}{2} m_2\omega^2 \left(y_j-L\right)^2 \right] + \nonumber \\
& & + 
\lambda_1 \sum_{1 \le j < k}^{N_1} \left(x_j-x_k\right)^2 + 
\lambda_2 \sum_{1 \le j < k}^{N_2} \left(y_j-y_k\right)^2 +  
\lambda_{12} \sum_{j=1}^{N_1} \sum_{k=1}^{N_2} \left(x_j-y_k\right)^2. \
\eeqn
Throughout this investigation we work in one spatial dimension and take $\hbar=1$.
There are $N_1$ bosons of type $1$
and $N_2$ bosons of type $2$,
the corresponding masses are $m_1$ and $m_2$,
the intra-species interaction strengths are $\lambda_1$ and $\lambda_2$,
and the inter-species interaction strength is $\lambda_{12}$.
The total number of bosons is denoted by $N=N_1+N_2$.

Expressed in terms of the Jacoby coordinates of the mixture,
$Q_k = \frac{1}{\sqrt{k(k+1)}} \sum_{j=1}^{k} (x_{k+1}-x_j), \ 1 \le k \le N_1-1$;
$Q_{N_1-1+k} = \frac{1}{\sqrt{k(k+1)}} \sum_{j=1}^{k} (y_{k+1}-y_j), \ 1 \le k \le N_2-1$;
$Q_{N-1} = \sqrt{N_1N_2}\left(X_{N_1}-Y_{N_2}\right)$;
and
$Q_N = \frac{m_1N_1}{M}X_{N_1}  + \frac{m_2N_2}{M}Y_{N_2}$,
where $X_{N_1}=\frac{1}{N_1}\sum_{j=1}^{N_1} x_j$
and $Y_{N_2}=\frac{1}{N_2}\sum_{j=1}^{N_2} y_j$ are the center-of-mass coordinates
of the individual species,
the Hamiltonian (\ref{HAM_MIX_CAR}) is diagonalized and takes on the form
\beqn\label{HAM_MIX_HAM}
& & \!\!\!\!\!\!\!\! \hat H(Q_1,\ldots,Q_N) =
\sum_{k=1}^{N_1-1} \left( -\frac{1}{2m_1} \frac{\partial^2}{\partial Q_k^2} + 
\frac{1}{2} m_1\Omega_1^2 Q_k^2 \right) + 
\sum_{k=N_1}^{N-2} \left( -\frac{1}{2m_2} \frac{\partial^2}{\partial Q_k^2} + 
\frac{1}{2} m_2\Omega_2^2 Q_k^2 \right) + \nonumber \\
& & \!\!\!\!\!\!\!\! + \left[-\frac{1}{2M_{12}} \frac{\partial^2}{\partial Q_{N-1}^2} + 
\frac{1}{2} M_{12} \Omega_{12}^2 \left(Q_{N-1}-Q^0_{N-1}\right)^2\right] 
+ \left[-\frac{1}{2M} \frac{\partial^2}{\partial Q_N^2} + 
\frac{1}{2} M \omega^2 \left(Q_N-Q^0_N\right)^2\right] + \nonumber \\
& & \!\!\!\!\!\!\!\! + \frac{1}{2} \frac{m_1N_1m_2N_2}{M} \omega^2 L^2 \left(1 - \frac{\omega^2}{\Omega_{12}^2}\right),
\eeqn
where $M_{12} = \frac{m_1m_2}{M}$ and $M=N_1m_1+N_2m_2$
are the relative center-of-mass and total masses, respectively.
The shifts in the relative center-of-mass and center-of-mass coordinates
\beqn\label{HAM_MIX_SHIFTS}
Q^0_{N-1} = - \sqrt{N_1N_2} \frac{\omega^2}{\Omega_{12}^2} L, \qquad
Q^0_N = \frac{m_2N_2}{M}L
\eeqn
emerge from completing the squares but actually govern,
as we shall discuss below, the demixing of the two condensates.
In particular, unless $L \ne 0$, demixing cannot occur in our model.
The specific case $L=0$, which cannot exhibit demixing,
has been analyzed in \cite{JPA_2017}.
Finally, the interaction-dressed frequencies are given by
\beqn\label{HAM_MIX_FREQ}
& & \Omega_{12} = \sqrt{\omega^2 + 2\left(\frac{N_1}{m_2}+\frac{N_2}{m_1}\right)\lambda_{12}}, \nonumber \\
& & \Omega_{1} = \sqrt{\omega^2 + \frac{2}{m_1}\left(N_1\lambda_1+N_2\lambda_{12}\right)}, \quad
\Omega_{2} = \sqrt{\omega^2 + \frac{2}{m_2}\left(N_2\lambda_2+N_1\lambda_{12}\right)}
\eeqn
and seen to be independent of $L$.
For attractive inter-species interaction $\frac{\Omega_{12}}{\omega} > 1$
and for repulsive interaction $0 < \frac{\Omega_{12}}{\omega} < 1$.
Correspondingly, $|Q^0_{N-1}| \to 0$ with increasing inter-species attraction
and $|Q^0_{N-1}| \to \infty$ with increasing repulsion.
From a different perspective,
for an attractive inter-species interaction the last term
in the Hamiltonian (\ref{HAM_MIX_HAM})
and hence the total energy, see below, decrease as $L \to 0$
whereas for repulsive inter-species interaction the total energy
decreases as $L \to \infty$,
as is expected for attractive and repulsive forces.
The center-of-mass of the mixture
is localized at $Q^0_{N-1}$ independently of any interaction.

With diagonalization of the Hamiltonian (\ref{HAM_MIX_CAR}) to (\ref{HAM_MIX_HAM}),
the wavefunction and energy of the ground state
are readily given by
\beqn\label{Psi_JAC_MIX}
& & \Psi(Q_1,\ldots,Q_N) = 
\left(\frac{m_1\Omega_1}{\pi}\right)^{\frac{N_1-1}{4}}
\left(\frac{m_2\Omega_2}{\pi}\right)^{\frac{N_2-1}{4}}
\left(\frac{M_{12}\Omega_{12}}{\pi}\right)^{\frac{1}{4}}
\left(\frac{M\omega}{\pi}\right)^{\frac{1}{4}}
\times \nonumber \\
& & \times
e^{-\frac{1}{2} \left\{ m_1\Omega_1 \sum_{k=1}^{N_1-1} Q_k^2 + m_2\Omega_2 \sum_{k=N_1}^{N-2} Q_k^2 
+ M_{12}\Omega_{12} \left[Q_{N-1}-Q^0_{N-1}\right]^2 + M\omega \left[Q_N-Q^0_N\right]^2 \right\}}  \
\eeqn
and
\beqn\label{E_MIX}
& & E = \frac{1}{2} \Bigg[(N_1-1) \sqrt{\omega^2 + \frac{2}{m_1}(N_1\lambda_1+N_2\lambda_{12})} +
(N_2-1)\sqrt{\omega^2 + \frac{2}{m_2}(N_2\lambda_2+N_1\lambda_{12})} + \nonumber \\ 
& & + \sqrt{\omega^2 + 2\left(\frac{N_1}{m_2}+\frac{N_2}{m_1}\right)\lambda_{12}} + \omega +
\frac{m_1N_1m_2N_2}{M} \omega^2 L^2 \left(1 - \frac{\omega^2}{\Omega_{12}^2}\right)\Bigg].
\eeqn
It is instructive to compare the structures of the wavefunction (\ref{Psi_JAC_MIX}) and energy (\ref{E_MIX}) to
the solution of the problem for $L=0$ \cite{JPA_2017}.
In particular, the center-of-mass and relative center-of-mass are not centered anymore at their origins
and a term proportional to $L^2$ is added to the energy.
We shall analyze the implications of these structures shortly.
 
To translate the wavefunction to the laboratory frame,
we plug in the mixture's Jacoby coordinates explicitly into (\ref{Psi_JAC_MIX}). 
Furthermore, the shifts of the relative center-of-mass and center-of-mass coordinates (\ref{HAM_MIX_SHIFTS}) 
have to be `translated' to corresponding coordinates' shifts $x_0$ and $y_0$ of the two species in the laboratory frame.
Using the inverse relations
$X_{N_1} = Q_N + \frac{m_2}{M}\sqrt{\frac{N_2}{N_1}} Q_{N-1}$
and
$Y_{N_2} = Q_N - \frac{m_1}{M}\sqrt{\frac{N_1}{N_2}} Q_{N-1}$
the final result for the wavefunction is given by
\beqn\label{LAB_SHIFT_PSI}
& & \Psi(x_1,\ldots,x_{N_1},y_1,\ldots,y_{N_2}) = 
\left(\frac{m_1\Omega_1}{\pi}\right)^{\frac{N_1-1}{4}}
\left(\frac{m_2\Omega_2}{\pi}\right)^{\frac{N_2-1}{4}}
\left(\frac{M_{12}\Omega_{12}}{\pi}\right)^{\frac{1}{4}}
\left(\frac{M\omega}{\pi}\right)^{\frac{1}{4}}
\times \nonumber \\
& & \times
e^{-\frac{\alpha_1}{2} \sum_{j=1}^{N_1}\left(x_j-x_0\right)^2 - 
\beta_1 \sum_{1 \le j < k}^{N_1} \left(x_j-x_0\right) \left(x_k-x_0\right)}
e^{-\frac{\alpha_1}{2} \sum_{j=1}^{N_2}\left(y_j-y_0\right)^2 - 
\beta_2 \sum_{1 \le j < k}^{N_2} \left(y_j-y_0\right) \left(y_k-y_0\right)} \times \nonumber \\
& & \times e^{+\gamma \sum_{j=1}^{N_1} \sum_{k=1}^{N_2} \left(x_j-x_0\right)\left(y_k-y_0\right)}, \
\eeqn
where the coefficients $\alpha$, $\beta$, and $\gamma$ are collected in \ref{APP_RDMs}.
The shifts of the coordinates in the laboratory frame are given by
the expectation values
of the individual species' center-of-mass operators
\beqn\label{LAB_SHIFT_x0_y0}
x_0 = \langle\Psi|\hat X_{N_1}|\Psi\rangle =
\frac{m_2N_2}{M}\left(1-\frac{\omega^2}{\Omega^2_{12}}\right)L, \quad
y_0 = \langle\Psi|\hat Y_{N_2}|\Psi\rangle =
\left[1 - \frac{m_1N_1}{M}\left(1-\frac{\omega^2}{\Omega^2_{12}}\right)\right]L.
\eeqn
We can now discuss the meaning of repulsion and attraction between the two
condensates which becomes transparent within our model.
When $\lambda_{12}=0$, species $1$ is localized at $x_0=0$ and species $2$ at $y_0=L$.
For repulsive inter-species interaction the two species grow apart: $x_0$ decreases and $y_0$ increases;
whereas for attractive inter-species interaction the two species come closer together: $x_0$ increases and $y_0$ decreases.
Side by side, 
the term added to the energy 
[last term in (\ref{E_MIX})] takes on an appealing form as a function of the 
individual species' center-of-mass expectation values (\ref{LAB_SHIFT_x0_y0}),
\beqn\label{ENERGY_x0_y0}
\frac{1}{2}\frac{m_1N_1m_2N_2}{M} \omega^2 L^2 \left(1 - \frac{\omega^2}{\Omega_{12}^2}\right) =
\frac{1}{2}m_1N_1 \omega^2 x_0^2 + \frac{1}{2}m_2N_2 \omega^2 (y_0-L)^2 + \lambda_{12}N_1N_2\left(x_0-y_0\right)^2,
\eeqn
also see \ref{GP_APP}.
The added energy term (\ref{ENERGY_x0_y0}) 
can be interpreted as
the `classical potential energy' of a $N_1$ particles of mass $m_1$
shifted by $x_0$ in a harmonic potential localized at the origin,
$N_2$ particles of mass $m_2$
shifted by $y_0-L$ in a harmonic potential localized at $L$,
and the corresponding energy of their mutual harmonic interaction
which scales like the square of their distance $(x_0-y_0)^2$.
We emphasis that all terms
of the mixture's energy (\ref{E_MIX}), except that originating from the center-of-mass, 
depend on the inter-species interaction
$\lambda_{12}$ and hence vary along the demixing pathway.

Let us proceed to many-body quantities derived from the many-particle
wavefunction of the mixture (\ref{LAB_SHIFT_PSI}).
The all-particle density matrix of the mixture is defined by (here normalized to unity):
\beqn\label{MB_DENS_SHIFTS}
& & \rho_{12}^{(N)}(x_1,\ldots,x_{N_1},y_1,\ldots,y_{N_2},x'_1,\ldots,x'_{N_1},y'_1,\ldots,y'_{N_2})  = \nonumber \\
& & = \left(\frac{m_1\Omega_1}{\pi}\right)^{\frac{N_1-1}{2}}
\left(\frac{m_2\Omega_2}{\pi}\right)^{\frac{N_2-1}{2}}
\left(\frac{M_{12}\Omega_{12}}{\pi}\right)^{\frac{1}{2}}
\left(\frac{M\omega}{\pi}\right)^{\frac{1}{2}} \times \nonumber \\
& & \times 
e^{-\frac{\alpha_1}{2} \sum_{j=1}^{N_1} \left[(x_j-x_0)^2+(x'_j-x_0)^2\right] - 
\beta_1 \sum_{1 \le j < k}^{N_1} \left[(x_j-x_0)(x_k-x_0) + (x'_j-x_0)(x'_k-x_0)\right]} \times \nonumber \\
& & \times 
e^{-\frac{\alpha_2}{2} \sum_{j=1}^{N_2} \left[(y_j-y_0)^2+(y'_j-y_0)^2\right] - 
\beta_2 \sum_{1 \le j < k}^{N_2} \left[(y_j-y_0)(y_k-y_0) + (y'_j-y_0)(y'_k-y_0)\right]} \times \nonumber \\
& & \times
e^{+\gamma \sum_{j=1}^{N_1} \sum_{k=1}^{N_2} \left[(x_j-x_0)(y_k-y_0) + (x'_j-x_0)(y'_k-y_0)\right]}.  \
\eeqn 
The integration of (\ref{MB_DENS_SHIFTS}) to the intra-species and inter-species
reduced density matrices \cite{RDM}
follows the lines of \cite{JPA_2017,Floquet_2020} and are not reproduced here.
The final results for
the lowest-order intra-species and inter-species reduced density matrices are given by
\beqn\label{RDMs_1_2_12}
& & \rho_1^{(1)}(x,x') = N_1 \left(\frac{\alpha_1+C_{1,0}}{\pi}\right)^\frac{1}{2}
e^{-\frac{\alpha_1}{2}\left[(x-x_0)^2+(x'-x_0)^2\right]} e^{-\frac{1}{4}C_{1,0}\left[(x-x_0)+(x'-x_0)\right]^2},
\nonumber \\
& & \rho_2^{(1)}(y,y') = N_2 \left(\frac{\alpha_1+C'_{0,1}}{\pi}\right)^\frac{1}{2}
e^{-\frac{\alpha_2}{2}\left[(y-y_0)^2+(y'-y_0)^2\right]} e^{-\frac{1}{4}C'_{0,1}\left[(y-y_0)+(y'-y_0)\right]^2},
\nonumber \\
& &
\rho_{12}^{(2)}(x,x',y,y') = N_1 N_2 
\left[\frac{(\alpha_1+C_{1,1})(\alpha_2+C'_{1,1})-D_{1,1}^2}{\pi^2}\right]^\frac{1}{2}
e^{-\frac{\alpha_1}{2}\left[(x-x_0)^2+(x'-x_0)^2\right]} \times \nonumber \\
& & \times e^{-\frac{\alpha_2}{2}\left[(y-y_0)^2+(y'-y_0)^2\right]} 
\times e^{-\frac{1}{4}C_{1,1}\left[(x-x_0)+(x'-x_0)\right]^2}
e^{-\frac{1}{4}C'_{1,1}\left[(y-y_0)+(y'-y_0)\right]^2} \times \nonumber \\
& & \times e^{+\frac{1}{2}D_{1,1}\left[(x-x_0)+(x'-x_0)\right]\left[(y-y_0)+(y'-y_0)\right]}
e^{+\frac{1}{2}D'_{1,1}\left[(x-x_0)-(x'-x_0)\right]\left[(y-y_0)-(y'-y_0)\right]}, \
\eeqn
where the various coefficients $C_{1,0}$, $C'_{0,1}$, $C_{1,1}$, $C'_{1,1}$, $D_{1,1}$, and $D'_{1,1}$
are results of coupled recursive relations \cite{JPA_2017} and prescribed in \ref{APP_RDMs}.
The respective densities, i.e., the diagonals of the reduced density matrices (\ref{RDMs_1_2_12}), are given by
\beqn\label{DENS_1_2_12}
& & \rho_1^{(1)}(x) = N_1 \left(\frac{\alpha_1+C_{1,0}}{\pi}\right)^\frac{1}{2}
e^{-(\alpha_1+C_{1,0})(x-x_0)^2},
\nonumber \\
& & \rho_2^{(1)}(y) = N_2 \left(\frac{\alpha_2+C'_{0,1}}{\pi}\right)^\frac{1}{2}
e^{-(\alpha_2+C'_{0,1})(y-y_0)^2},
\nonumber \\
& &
\rho_{12}^{(2)}(x,y) = N_1 N_2 
\left[\frac{(\alpha_1+C_{1,1})(\alpha_2+C'_{1,1})-D_{1,1}^2}{\pi^2}\right]^\frac{1}{2}
e^{-(\alpha_1+C_{1,1})(x-x_0)^2} \times \nonumber \\
& & \times e^{-(\alpha_2+C'_{1,1})(y-y_0)^2} e^{+2D_{1,1}(x-x_0)(y-y_0)}. \
\eeqn
Clearly, the intra-species densities are localized around $x_0$ or $y_0$
whereas the inter-species density is localized
both at $x_0$ and $y_0$.
Since all these many-body quantities are given analytically,
we can evaluate them explicitly
at each point along the demixing pathway.

The extent to which the two condensates mix along the demixing pathway
can be quantified using the spatial overlap between their corresponding densities (\ref{DENS_1_2_12}).
Explicitly, we consider the overlap of the square roots of the one-particle densities per particle,
\beqn\label{OVERLAP_12}
& & S_{12} = \int_{-\infty}^{+\infty} dx \sqrt{\frac{\rho_1^{(1)}(x)}{N_1}} \sqrt{\frac{\rho_2^{(1)}(x)}{N_2}} = \nonumber \\
& & =
\left(\frac{\sqrt{(\alpha_1+C_{1,0})(\alpha_2+C'_{0,1})}}{\frac{1}{2}[(\alpha_1+C_{1,0})+(\alpha_2+C'_{0,1})]}\right)^\frac{1}{2}
e^{-\frac{(\alpha_1+C_{1,0})(\alpha_2+C'_{0,1})}{2[(\alpha_1+C_{1,0})+(\alpha_2+C'_{0,1})]}(x_0-y_0)^2} =
\nonumber \\
& & =
\left(\frac{\sqrt{(\alpha_1+C_{1,0})(\alpha_2+C'_{0,1})}}{\frac{1}{2}[(\alpha_1+C_{1,0})+(\alpha_2+C'_{0,1})]}\right)^\frac{1}{2}
e^{-\frac{(\alpha_1+C_{1,0})(\alpha_2+C'_{0,1})}{2[(\alpha_1+C_{1,0})+(\alpha_2+C'_{0,1})]}\frac{\omega^4}{\Omega_{12}^4}L^2}. \
\eeqn
Definition (\ref{OVERLAP_12}) seems natural since
it reduces to unity when the two densities are equal.
Generally, when the distance between the two condensates $y_0-x_0 = \frac{\omega^2}{\Omega^2_{12}}L$ increases, i.e., for $\frac{\Omega_{12}}{\omega}<1$,
their overlap decreases and vice versa.
Of course, expression (\ref{OVERLAP_12}) gives the precise value of the two condensates' overlap
as a function of all parameters; the masses, interaction strengths, and the number of particles of each species.
Below, we obtain the analogous expression evaluated at the mean-field level of theory,
in which the interplay between the intra-species and inter-species interactions can
be straightforwardly analyzed.

Let us move to the mean-field solution for demixing which is obtained analytically as follows.
The derivation generalizes that in \cite{JPA_2017,Floquet_2020,Cohen_1985}.
The Gross-Pitaevskii ansatz for the mixture's wavefunction is
\beqn\label{MF_MIX_ANZ}
& & 
\Phi(x_1,\ldots,x_{N_1},y_1,\ldots,y_{N_2}) = \prod_{j=1}^{N_1} \phi_1(x_j) \prod_{k=1}^{N_2} \phi_2(y_k).
\eeqn
The orbitals $\phi_1(x)$ and $\phi_2(y)$ have to be determined self consistently.
Sandwiching the many-body Hamiltonian (\ref{HAM_MIX_CAR})
with the mean-field ansatz (\ref{MF_MIX_ANZ}) one gets the  
Gross-Pitaevskii energy functional.
Minimizing the latter with respect to the shapes
of the two normalized orbitals
$\phi_1(x)$ and $\phi_2(y)$,
one obtains the coupled nonlinear integro-differential equations
\beqn\label{MIX_EQ_TDGP_LAB}
& &
\Bigg\{-\frac{1}{2m_1} \frac{\partial^2}{\partial x^2} + \frac{1}{2} m_1 \omega^2 x^2
+ \Lambda_1 \int dx' |\phi_1(x')|^2 (x-x')^2 + \nonumber \\ & & 
+ \Lambda_{21} \int dy |\phi_2(y)|^2 (x-y)^2 \Bigg\} \phi_1(x) = 
\mu_1 \phi_1(x),
\nonumber \\
& & 
\Bigg\{-\frac{1}{2m_2} \frac{\partial^2}{\partial y^2} + \frac{1}{2} m_2 \omega^2 (y-L)^2 
+ \Lambda_2 \int dy' |\phi_2(y')|^2 (y-y')^2 + \nonumber \\ & & 
+ \Lambda_{12} \int dx |\phi_1(x)|^2 (x-y)^2 \Bigg\} \phi_2(y) = 
\mu_2 \phi_2(y), \
\eeqn
where $\mu_1$ and $\mu_2$ stand for the respective chemical potentials
and the mean-field interaction parameters are given by 
$\Lambda_1=\lambda_1(N_1-1)$,
$\Lambda_2=\lambda_1(N_2-1)$,
$\Lambda_{12}=\lambda_{12}N_1$, and $\Lambda_{21}=\lambda_{12}N_2$.
Recall that within the
Gross-Pitaevskii treatment of demixing
only the interaction parameters $\Lambda_1$, $\Lambda_2$, $\Lambda_{12}$, and $\Lambda_{21}$ 
appear. 

The coupled Gross-Pitaevskii equations (\ref{MIX_EQ_TDGP_LAB}) admit an analytic solution.
This is intriguing in itself,
as we are not aware of other analytical mean-field solutions
for demixing of two spatially-inhomogeneous
Bose-Einstein condensates. 
The final result for the orbitals reads
\beqn\label{MIX_GP_OR_SHIFT}
& & \phi_1(x) = \left(\frac{m_1}{\pi}\sqrt{\omega^2 + \frac{2}{m_1}(\Lambda_1+\Lambda_{21})}\right)^{\frac{1}{4}}
e^{-\frac{m_1}{2}\sqrt{\omega^2 + \frac{2}{m_1}(\Lambda_1+\Lambda_{21})} (x-x_0)^2}, \nonumber \\
& & \phi_2(y) = \left(\frac{m_2}{\pi}\sqrt{\omega^2 + \frac{2}{m_2}(\Lambda_2+\Lambda_{12})}\right)^{\frac{1}{4}}
e^{-\frac{m_2}{2}\sqrt{\omega^2 + \frac{2}{m_2}(\Lambda_2+\Lambda_{12})} (y-y_0)^2}, \
\eeqn
where further details
are collected in \ref{GP_APP}.
Finally, the Gross-Pitaevskii energy of the mixture 
takes on the following form,
expressed as a function of the interaction parameters only:
\beqn\label{MIX_GP_E}
& & \varepsilon^{GP} = \frac{E^{GP}}{N} =
\frac{1}{2}\frac{\Lambda_{12}\sqrt{\omega^2 + \frac{2}{m_1}(\Lambda_1+\Lambda_{21})} +
\Lambda_{21}\sqrt{\omega^2 + \frac{2}{m_2}(\Lambda_2+\Lambda_{12})}}{\Lambda_{12}+\Lambda_{21}} +
\nonumber \\
& & +
\frac{1}{2}\frac{\Lambda_{12}m_1\omega^2x_0^2+\Lambda_{21}m_2\omega^2(y_0-L)^2}{\Lambda_{12}+\Lambda_{21}} +
\frac{\Lambda_{12}\Lambda_{21}(x_0-y_0)^2}{\Lambda_{12}+\Lambda_{21}}. \
\eeqn
Indeed, the first line in (\ref{MIX_GP_E}) is the mean-field energy as if the two harmonic traps overlap ($L=0$), see \cite{JPA_2017}, and the second line
is precisely the `potential-energy-and-interaction' term
added at the many-body level of theory to describe the demixing for $L \ne 0$,
see (\ref{E_MIX}) and (\ref{ENERGY_x0_y0}).

Let us intermediately summarize.
We have put forward an exactly-solvable model for demixing of two Bose-Einstein condensates,
whose many-body and mean-field ground-state solutions are given in closed and analytical forms.
We can now ask further questions, on energetics, condensation, correlations, and on other properties,
first at the many-body level of theory and than at the mean-field level of theory,
and investigate the respective differences.
From what we have depicted so far above,
the model (\ref{HAM_MIX_CAR}) looks sufficiently rich
such that a detailed account can only find sufficient room beyond the present paper.
We hence proceed with exploration of just two additional selected quantities.   

We return to the degree of mixing of the two condensates
characterized by their spatial overlap (\ref{OVERLAP_12}),
but now at the mean-field level.
The Gross-Pitaevskii densities per particle are nothing but $\phi^2_1(x)$ and $\phi^2_2(y)$
[the orbitals (\ref{MIX_GP_OR_SHIFT}) are real-valued functions].
Consequently, we readily find
\beqn\label{OVERLAP_12_GAP}
& & \!\!\!\!\!\!\!\!\!\!\!\! S^{GP}_{12} = \int_{-\infty}^{+\infty} dx \phi_1(x)\phi_2(x) =
\left(\frac{\sqrt{m_1m_2\sqrt{\omega^2 + \frac{2}{m_1}(\Lambda_1+\Lambda_{21})}\sqrt{\omega^2 + \frac{2}{m_2}(\Lambda_2+\Lambda_{12})}}}{\frac{1}{2}\left[m_1\sqrt{\omega^2 + \frac{2}{m_1}(\Lambda_1+\Lambda_{21})}+m_2\sqrt{\omega^2 + \frac{2}{m_2}(\Lambda_2+\Lambda_{12})}\right]}\right)^\frac{1}{2} \times \nonumber \\
& & \!\!\!\!\!\!\!\!\!\!\!\! \times 
e^{-\frac{m_1m_2\sqrt{\omega^2 + \frac{2}{m_1}(\Lambda_1+\Lambda_{21})}\sqrt{\omega^2 + \frac{2}{m_2}(\Lambda_2+\Lambda_{12})}}{2\left[m_1\sqrt{\omega^2 + \frac{2}{m_1}(\Lambda_1+\Lambda_{21})}+m_2\sqrt{\omega^2 + \frac{2}{m_2}(\Lambda_2+\Lambda_{12})}\right]}(x_0-y_0)^2},\
\eeqn
where $x_0-y_0= - \frac{\omega^2}{\Omega^2_{12}}L$ just like in the many-body case,
see \ref{GP_APP}.
Equation (\ref{OVERLAP_12_GAP}) shows in a transparent manner the dependence of the spatial overlap between
the two condensates on the masses, $m_1$ and $m_2$, and intra-species $\Lambda_1$, $\Lambda_2$ and inter-species
$\Lambda_{12}$, $\Lambda_{21}$ interaction parameters.
One can push the analysis further in the case the parameters of both species are equal, i.e.,
$m_2=m_1$, $\Lambda_2=\Lambda_1$, and $\Lambda_{12}=\Lambda_{21}$.
Then,
$S^{GP}_{12}=e^{-\frac{m_1}{4}\sqrt{\omega^2 + \frac{2}{m_1}(\Lambda_1+\Lambda_{21})}\frac{\omega^4}{\Omega^4_{12}}L^2}$.
For a given intra-species interaction $\Lambda_1$, attraction or repulsion,
the overlap decreases monotonously with inter-species repulsion and increases monotonously with inter-species attraction.
Furthermore,
for a given inter-species interaction $\Lambda_{21}$, attraction or repulsion,
the overlap increases monotonously with intra-species repulsion and decreases monotonously with intra-species attraction.
The latter reflects the common wisdom that it is more difficult to
spatially-separate condensates when they are made of repulsive species.

Perhaps, the most obvious difference between the many-body and mean-field wavefunctions of the mixture,
equations (\ref{LAB_SHIFT_PSI}) and (\ref{MF_MIX_ANZ}), respectively,
is that in the many-body treatment the two species are entangled whereas 
in the mean-field description,
using the separable product state,
the two species are obviously not entangled.
Thus, finally, we move to the Schmidt decomposition of the many-body wavefunction,
thereby generalizing recent results in the specific case of a symmetric mixture \cite{Atoms_2021}.
We begin from and employ Meher's formula which can be written as follows:
\beqn\label{MEHLER_XY}
& &\left(\frac{s}{\pi}\right)^{\frac{1}{2}}
e^{-\frac{1}{2}\frac{(1+\rho^2)s}{1-\rho^2}\left(x^2+y^2\right)} \,
e^{+\frac{2\rho s}{1-\rho^2}xy}
= \nonumber \\
& & 
\quad = \sum_{n=0}^\infty \sqrt{1-\rho^2}\rho^n \frac{1}{\sqrt{2^n n!}}
\left(\frac{s}{\pi}\right)^{\frac{1}{4}} H_n(\sqrt{s}x) e^{-\frac{1}{2}s x^2}
\frac{1}{\sqrt{2^n n!}} \left(\frac{s}{\pi}\right)^{\frac{1}{4}} H_n(\sqrt{s}y) e^{-\frac{1}{2}s {y}^2}, \
\eeqn
where the parameters $s>0$, $1 > \rho \ge 0$
for Schmidt decomposition of the wavefunction,
and $H_n$ are the Hermite polynomials.
The wavefunction (\ref{LAB_SHIFT_PSI})
is rewritten in terms of the Jacoby coordinates of each of the species,
including the shifts of the coordinates $x_0$ and $y_0$,
\beqn\label{PSI_4_MEHLER}
& & \Psi(\bar X_1,\ldots,\bar X_{N_1},\bar Y_1,\ldots,\bar Y_{N_2}) = 
\left(\frac{m_1\Omega_1}{\pi}\right)^{\frac{N_1-1}{4}}
\left(\frac{m_2\Omega_2}{\pi}\right)^{\frac{N_2-1}{4}}
\left(\frac{M_{12}\Omega_{12}}{\pi}\right)^{\frac{1}{4}}
\left(\frac{M\omega}{\pi}\right)^{\frac{1}{4}}
\times \nonumber \\
& & \times
e^{-\frac{1}{2}\left(m_1\Omega_1 \sum_{k=1}^{N_1-1} \bar X_k^2 + m_2\Omega_2 \sum_{k=1}^{N_2-1} \bar Y_k^2\right)} \times
\nonumber \\
& & \times e^{-\frac{1}{2}\left(m_1 \frac{m_2N_2\Omega_{12} +m_1N_1 \omega}{M} \bar X^2_{N_1} +
m_2 \frac{m_1N_1\Omega_{12}+m_2N_2\omega}{M} \bar Y^2_{N_2}\right)}
e^{+\frac{m_1m_2\sqrt{N_1N_2}}{M}(\Omega_{12}-\omega) \bar X_{N_1}\bar Y_{N_2}},  \
\eeqn 
where we denote for brevity here and hereafter
$\bar X_k = \frac{1}{\sqrt{k(k+1)}} \sum_{j=1}^{k} [(x_{k+1}-x_0)-(x_j-x_0)], \ 1 \le k \le N_1-1$;
$\bar Y_k = \frac{1}{\sqrt{k(k+1)}} \sum_{j=1}^{k} [(y_{k+1}-y_0)-(y_j-y_0)], \ 1 \le k \le N_2-1$;
$\bar X_{N_1}=\frac{1}{N_1}\sum_{j=1}^{N_1} (x_j-x_0)$; and 
$\bar Y_{N_2}=\frac{1}{N_2}\sum_{j=1}^{N_2} (y_j-y_0)$.
Furthermore, in (\ref{PSI_4_MEHLER}) it is convenient to treat first the case of mixing,
i.e., of attractive inter-species interaction $\Omega_{12} > \omega$;
the slight modifications in the treatment of demixing,
i.e., for repulsive inter-species interaction $\Omega_{12} < \omega$,
are put forward below.  
Clearly, the wavefunction (\ref{PSI_4_MEHLER}) boils down to that of
the symmetric mixture when the parameters of species $1$ and species $2$ bosons are equal and for $L=0$.
On the other hand, unlike the Schmidt decomposition of the symmetric mixture \cite{Atoms_2021} and
before Mehler's formula (\ref{MEHLER_XY}) can be applied,
equation (\ref{PSI_4_MEHLER}) would require a squeeze mapping of $\bar X_{N_1}$ and $\bar Y_{N_2}$.
Thus, defining
\beqn\label{SQUEEZE}
\widetilde X^2_{N_1} \equiv \bar X^2_{N_1} 
\sqrt{\frac{m_1\left(m_2N_2\Omega_{12} +m_1N_1 \omega\right)}
{m_2\left(m_1N_1\Omega_{12}+m_2N_2\omega\right)}}, \quad
\widetilde Y^2_{N_2} \equiv \bar Y^2_{N_2} 
\sqrt{\frac{m_2\left(m_1N_1\Omega_{12}+m_2N_2\omega\right)}
{m_1\left(m_2N_2\Omega_{12} +m_1N_1 \omega\right)}}
\eeqn
(satisfying $\widetilde X_{N_1}\widetilde Y_{N_2} = \bar X_{N_1} \bar Y_{N_2}$),
the last row of the wavefunction (\ref{PSI_4_MEHLER}) transforms and reads
\beqn\label{PSI_4_MEHLER_SQUEEZE}
& & e^{-\frac{1}{2} \frac{m_1m_2\sqrt{N_1N_2}}{M}
\sqrt{(\Omega_{12}+\omega)^2 + \frac{(m_1N_1-m_2N_2)^2}{m_1N_1m_2N_2} \Omega_{12}\omega}
\left(\widetilde X^2_{N_1} + \widetilde Y^2_{N_2}\right)}
e^{+\frac{m_1m_2\sqrt{N_1N_2}}{M}(\Omega_{12}-\omega) \widetilde X_{N_1} \widetilde Y_{N_2}}.  \
\eeqn 
Now, with equation (\ref{PSI_4_MEHLER_SQUEEZE}),
Mehler's formula can be directly applied.
The final result for the Schmidt decomposition reads
\beqn\label{WF_MIX_SCHMIDT}
& & \!\!\!\!\!\!\!\!\!\!\!\!
\Psi(\bar X_1,\ldots,\widetilde X_{N_1},\bar Y_1,\ldots,\widetilde Y_{N_2})
= \sum_{n=0}^{\infty} \sqrt{1-\rho^2} \rho^n \Phi_{1,n}(\bar X_1,\ldots,\widetilde X_{N_1})
\Phi_{2,n}(\bar Y_1,\ldots,\widetilde Y_{N_2}), \nonumber \\
& & \!\!\!\!\!\!\!\!\!\!\!\!
\Phi_{1,n}(\bar X_1,\ldots,\widetilde X_{N_1}) = \left(\frac{m_1\Omega_1}{\pi}\right)^{\frac{N_1-1}{4}}
e^{-\frac{1}{2} m_1 \Omega_1 \sum_{k=1}^{N_1-1} \bar X_k^2}
\frac{1}{\sqrt{2^n n!}} \left(\frac{s}{\pi}\right)^{\frac{1}{4}} H_n\left(\sqrt{s}\widetilde X_{N_1}\right)
e^{-\frac{1}{2}s \widetilde X_{N_1}^2}, \nonumber \\
& & \!\!\!\!\!\!\!\!\!\!\!\!
\Phi_{2,n}(\bar Y_1,\ldots,\widetilde Y_{N_2}) = \left(\frac{m_2\Omega_2}{\pi}\right)^{\frac{N_2-1}{4}}
e^{-\frac{1}{2} m_2 \Omega_2 \sum_{k=1}^{N_2-1} \bar Y_k^2}
\frac{1}{\sqrt{2^n n!}} \left(\frac{s}{\pi}\right)^{\frac{1}{4}} H_n\left(\sqrt{s}\widetilde Y_{N_2}\right) e^{-\frac{1}{2}s \widetilde Y_{N_2}^2}, \
\eeqn
where the Schmidt parameters are
\beqn\label{WF_MIX_SCHMIDT_PAR}
& & \!\!\!\!\!\!\!\!\!\!\!\!\!\!\!\!\!\!\!\!\!\!\!\!
\rho =
\frac{\left[\frac{\sqrt{\left(\frac{\Omega_{12}}{\omega}+1\right)^2 + \frac{(m_1N_1-m_2N_2)^2}{m_1N_1m_2N_2} \frac{\Omega_{12}}{\omega}}+\left(\frac{\Omega_{12}}{\omega}-1\right)}{\sqrt{\left(\frac{\Omega_{12}}{\omega}+1\right)^2 + \frac{(m_1N_1-m_2N_2)^2}{m_1N_1m_2N_2} \frac{\Omega_{12}}{\omega}}-\left(\frac{\Omega_{12}}{\omega}-1\right)}\right]^{+\frac{1}{2}}-1}
{\left[\frac{\sqrt{\left(\frac{\Omega_{12}}{\omega}+1\right)^2 + \frac{(m_1N_1-m_2N_2)^2}{m_1N_1m_2N_2} \frac{\Omega_{12}}{\omega}}+\left(\frac{\Omega_{12}}{\omega}-1\right)}{\sqrt{\left(\frac{\Omega_{12}}{\omega}+1\right)^2 + \frac{(m_1N_1-m_2N_2)^2}{m_1N_1m_2N_2} \frac{\Omega_{12}}{\omega}}-\left(\frac{\Omega_{12}}{\omega}-1\right)}\right]^{+\frac{1}{2}}+1}, \qquad
s = \sqrt{m_1m_2\omega\Omega_{12}}.
\
\eeqn
Finally, the case of repulsive inter-species interaction $\Omega_{12} < \omega$
implies the assignments of, e.g., $\bar Y_{N_2} \to -\bar Y_{N_2}$
in (\ref{PSI_4_MEHLER}) and
$\rho \to -\rho$ in (\ref{WF_MIX_SCHMIDT_PAR}), similarly to \cite{Atoms_2021}.

Equation (\ref{WF_MIX_SCHMIDT_PAR}) quantifies precisely
as a function of the mixture's parameters the entanglement between
the two species along the demixing pathway.
It is instrumental to discuss a few limiting cases.
Without inter-species interaction, i.e., when $\Omega_{12}=\omega$,
one has $\rho=0$ and the two species are, of course, not entangled;
The condensates
themselves can possess strong intra-species interactions though.
When $\Omega_{12}$ is very large (strong inter-species attraction)
or very small (strong inter-species repulsion),
$\rho$ increases more and more towards unity
and a high degree of entanglement emerges.
Hence, within our model, a high degree of entanglement accompanies both mixing and demixing.
Last but not least, when a very large asymmetry between the two species exists,
explicitly, say, $m_1N_1 \gg m_2N_2$,
$\rho$ decreases more and more towards zero,
implying the entanglement diminishes further and further,
see in this context the situation for $L=0$ \cite{EPJD_2014}. 
As can be expected,
for symmetric mixtures and $L=0$ some of the above-obtained
results boil down to those in \cite{Atoms_2021}.
This is a suitable place to stop the current investigation.

\section{Concluding Remarks}\label{SUM_OUT}

In the present work
a solvable model mimicking demixing of two Bose-Einstein condensates
at the many-body level of theory is devised and investigated.
The wavefunction, energy, reduced density matrices, densities, spatial overlap,
and entanglement between the two condensates expressed as the 
Schmidt decomposition of the many-particle wavefunction
are given in closed form along the demixing pathway.
The connection between the center-of-mass and in particular relative center-of-mass coordinates
and demixing is elucidated, within our model. 
Furthermore,
the model is also solved analytically at the mean-field level of theory,
and the above-computed properties are expressed in closed form also
at the mean-field level of theory.
A short discussion on the differences between properties 
computed at the many-body and mean-field levels
of theory along the demixing pathway is made.

There are several research directions the present investigation can lead to,
of which we list the following three.
An immediate study would be a comprehensive comparison between many-body and
mean-field descriptions of demixing at the limit of an infinite number of particles.
Therein, some properties, like the energy per particle and densities per particle, would exactly coincide and other properties,
like variances per particle of many-particle observables and the overlap between the many-body
and mean-field wavefunctions, can differ substantially
[53-65].
For finite mixtures, the fragmentation \cite{FRAG} along the demixing pathway would be instrumental to follow.
Another research venue that is worth pursuing is benchmarking multiconfigurational
time-dependent Hartree methods
[67-70]
and other numerical approaches for bosonic mixtures along the demixing pathway.
Finally, a more distant but rewarding challenge would be the emulation and subsequent investigation of many-body
effects when scattering attractive bosonic clouds from a potential barrier, or off each other, fully analytically
[71-74].

\ack

This research was supported by the Israel Science Foundation (Grant No. 1516/19). 
We thank Alexej I. Streltsov for motivating discussions.

\appendix

\section{Coefficients of the inter-species and intra-species reduced density matrices}\label{APP_RDMs}

It can be shown that
the parameters of the wavefunction (\ref{LAB_SHIFT_PSI}) in the laboratory frame entering the reduced density matrices
are those of the wavefunction without the coordinates' shifts $x_0$ and $y_0$,
and therefore are given by \cite{JPA_2017}:
\beqn\label{WF_PAR}
& & \alpha_1 = m_1\Omega_1+\beta_1, \qquad 
\beta_1 = m_1 \left[- \Omega_1 \frac{1}{N_1} 
+ (m_2N_2\Omega_{12} + m_1N_1\omega) \frac{1}{MN_1} \right], \nonumber \\
& & \alpha_2 = m_2\Omega_2+\beta_2, \qquad
\beta_2 = m_2 \left[- \Omega_2 \frac{1}{N_2} 
+ (m_1N_1\Omega_{12} + m_2N_2\omega) \frac{1}{MN_2} \right], \nonumber \\
& & \gamma = \frac{m_1m_2}{M}(\Omega_{12}-\omega). \
\eeqn
Correspondingly, since the various integrations of the all-particle density matrix
(\ref{MB_DENS_SHIFTS}) 
are taken along the variables $(x'_j-x_0)=(x_j-x_0)$ and $(y'_k-y_0)=(y_k-y_0)$,
and hence, upon these integrations, the coordinates' shifts $x_0$ and $y_0$ can be eliminated,
the coefficients of the inter-species reduced density matrix (\ref{RDMs_1_2_12}) are also
those of the corresponding reduced density matrix without coordinates' shifts
\cite{JPA_2017}:
\beqn\label{COEFF_12}
& & \!\!\!\!\!\!\!\!\!\!
\alpha_1 + C_{1,1} = (\alpha_1-\beta_1)
\frac{[(\alpha_1-\beta_1)+N_1\beta_1][(\alpha_2-\beta_2)+(N_2-1)\beta_2]-\gamma^2N_1(N_2-1)}
{[(\alpha_1-\beta_1)+(N_1-1)\beta_1][(\alpha_2-\beta_2)+(N_2-1)\beta_2]-\gamma^2(N_1-1)(N_2-1)},
\nonumber \\
& & \!\!\!\!\!\!\!\!\!\!
\alpha_2 + C'_{1,1} = (\alpha_2-\beta_2)
\frac{[(\alpha_2-\beta_2)+N_2\beta_2][(\alpha_1-\beta_1)+(N_1-1)\beta_1]-\gamma^2N_2(N_1-1)}
{[(\alpha_1-\beta_1)+(N_1-1)\beta_1][(\alpha_2-\beta_2)+(N_2-1)\beta_2]-\gamma^2(N_1-1)(N_2-1)},
\nonumber \\
& & \!\!\!\!\!\!\!\!\!\!
D_{1,1} = \gamma \frac{(\alpha_1-\beta_1)(\alpha_2-\beta_2)}
{[(\alpha_1-\beta_1)+(N_1-1)\beta_1][(\alpha_2-\beta_2)+(N_2-1)\beta_2]-\gamma^2(N_1-1)(N_2-1)}, \nonumber \\
& & \!\!\!\!\!\!\!\!\!\!
D'_{1,1} = \gamma.
\eeqn
Analogously, the coefficients of the intra-species reduced density matrices (\ref{RDMs_1_2_12}) are given by \cite{JPA_2017}:
\beqn\label{COEFF_1_2}
& & \!\!\!\!\!\!\!\!\!\!
\alpha_1 + C_{1,0} = (\alpha_1-\beta_1)
\frac{[(\alpha_1-\beta_1)+N_1\beta_1][(\alpha_2-\beta_2)+N_2\beta_2]-\gamma^2N_1N_2}
{[(\alpha_1-\beta_1)+(N_1-1)\beta_1][(\alpha_2-\beta_2)+N_2\beta_2]-\gamma^2(N_1-1)N_2},
\nonumber \\
& & \!\!\!\!\!\!\!\!\!\!
\alpha_2 + C'_{0,1} = (\alpha_2-\beta_2)
\frac{[(\alpha_1-\beta_1)+N_1\beta_1][(\alpha_2-\beta_2)+N_2\beta_2]-\gamma^2N_1N_2}
{[(\alpha_2-\beta_2)+(N_2-1)\beta_2][(\alpha_1-\beta_1)+N_1\beta_1]-\gamma^2(N_2-1)N_1}.
\nonumber \\ \
\eeqn
Of course, all these coefficients depend
explicitly on the masses $m_1$, $m_2$,
interaction strengths $\lambda_1$, $\lambda_2$, $\lambda_{12}$,
and the numbers of particles $N_1$, $N_2$,
and vary along the demixing pathway.

\section{Further details of the solution of the coupled Gross-Pitaevskii equations
and the coordinates' shifts}\label{GP_APP}

It is useful to rewrite (\ref{MIX_EQ_TDGP_LAB}) in terms of shifts of the coordinates $x_0$ and $y_0$,
which themselves have to be determined explicitly within mean-field theory.
Making use of the normalization of $\phi_1(x)$ and $\phi_2(y)$ 
and that they are even functions with respect to $x_0$ and $y_0$
[see the obtained self-consistent solution (\ref{MIX_GP_OR_SHIFT}) given in the main text],
one finds
\beqn\label{MIX_EQ_TDGP_SHIFT}
& & \!\!\!\!\!\!\!\!\!\!\!\!
\Bigg\{-\frac{1}{2m_1} \frac{\partial^2}{\partial x^2} + \frac{1}{2} m_1 \omega^2 (x-x_0)^2
+ \Lambda_1 \int dx' |\phi_1(x')|^2 \left[(x-x_0)-(x'-x_0)\right]^2 + \nonumber \\
& & \!\!\!\!\!\!\!\!\!\!\!\!
+ \Lambda_{21} \int dy |\phi_2(y)|^2 \left[(x-x_0)-(y-y_0)\right]^2 \Bigg\} \phi_1(x) = 
\left[\mu_1 - \frac{1}{2}m_1x_0^2 - \Lambda_{21}(x_0-y_0)^2\right] \phi_1(x),
\nonumber 
\\
& & \!\!\!\!\!\!\!\!\!\!\!\!
\Bigg\{-\frac{1}{2m_2} \frac{\partial^2}{\partial y^2} + \frac{1}{2} m_2 \omega^2 (y-y_0)^2 
+ \Lambda_2 \int dy' |\phi_2(y')|^2 \left[(y-y_0)-(y'-y_0)\right]^2 + \nonumber \\
& & \!\!\!\!\!\!\!\!\!\!\!\!
+ \Lambda_{12} \int dx |\phi_1(x)|^2 \left[(x-x_0)-(y-y_0)\right]^2 \Bigg\} \phi_2(y) = 
\left[\mu_2 - \frac{1}{2}m_2(y_0-L)^2 - \Lambda_{12}(x_0-y_0)^2\right] \phi_2(y), \nonumber \\
& & \!\!\!\!\!\!\!\!\!\!\!\! \
\eeqn
where,
for the terms linear in $(x-x_0)$ and $(y-y_0)$ to drop out,
$x_0$ and $y_0$ must obey
\beqn\label{MF_LIN_x0_y0}
& & m_1 \omega^2 x_0 + 2\Lambda_{21}(x_0-y_0) = 0, \nonumber \\
& & m_2 \omega^2 (y_0-L) - 2\Lambda_{12}(x_0-y_0) = 0. \
\eeqn
The solution of the linear system (\ref{MF_LIN_x0_y0}) 
is $x_0=\frac{2\Lambda_{21}}{m_1 \Omega^2_{12}} L$
and $y_0=\left(1-\frac{2\Lambda_{12}}{m_2 \Omega^2_{12}}\right) L$, where
$\Omega_{12} = \sqrt{\omega^2 + 2\left(\frac{\Lambda_{12}}{m_2}+\frac{\Lambda_{21}}{m_1}\right)}$,
hence $x_0-y_0 = \left[2\left(\frac{\Lambda_{12}}{m_2}+\frac{\Lambda_{21}}{m_1}\right)\frac{1}{\Omega^2_{12}} - 1\right]L$.
These are exactly the same values found within the many-body solution, see (\ref{LAB_SHIFT_x0_y0}).
Intriguingly, the frequency of the relative center-of-mass Jacoby coordinate
in the many-body treatment is obtained (in the mean-field treatment) 
from re-expressing the mean-field equations using the coordinates' shifts $x_0$ and $y_0$.

Consequently, the Gross-Pitaevskii solution for the demixing scenario ($L\ne 0$)
can be related to the solution of the coupled equations without the shifts of coordinates ($L=0$).
The self-consistent orbitals are given in the main text, see (\ref{MIX_GP_OR_SHIFT}),
and the respective chemical potentials read
\beqn\label{MIX_GP_MU_1_2}
& & \mu_1 = \frac{1}{2}\left(\sqrt{\omega^2 + \frac{2}{m_1}(\Lambda_1+\Lambda_{21})} 
+ \frac{\Lambda_1}{\sqrt{\omega^2 + \frac{2}{m_1}(\Lambda_1+\Lambda_{21})}} 
+ \frac{\Lambda_{21}}{\sqrt{\omega^2 + \frac{2}{m_2}(\Lambda_2+\Lambda_{12})}}\right) + \nonumber \\
& & + \frac{1}{2}m_1x_0^2 + \Lambda_{21}(x_0-y_0)^2, \nonumber \\
& & \mu_2 = \frac{1}{2}\left(\sqrt{\omega^2 + \frac{2}{m_2}(\Lambda_2+\Lambda_{12})}
+ \frac{\Lambda_2}{\sqrt{\omega^2 + \frac{2}{m_2}(\Lambda_2+\Lambda_{12})}} 
+ \frac{\Lambda_{12}}{\sqrt{\omega^2 + \frac{2}{m_1}(\Lambda_1+\Lambda_{21})}}\right) + \nonumber \\
& & + \frac{1}{2}m_2(y_0-L)^2 + \Lambda_{12}(x_0-y_0)^2. \
\eeqn
The total Gross-Pitaevskii energy (\ref{MIX_GP_E}) for demixing 
is equivalently obtained from the chemical potentials and the interaction energy,
$E^{GP} = N_1\mu_1+N_2\mu_2 - \frac{N_1}{2}\Big[\Lambda_1\int dx dx' |\phi_1(x)|^2 |\phi_1(x')|^2 (x-x')^2 +
\Lambda_{21}\int dx dy |\phi_1(x)|^2 |\phi_2(y)|^2 (x-y)^2\Big]
- \frac{N_2}{2}\Big[\Lambda_2\int dy dy' |\phi_2(y)|^2 |\phi_2(y')|^2 (y-y')^2 +\break\hfill 
\Lambda_{12}\int dx dy |\phi_1(x)|^2 |\phi_2(y)|^2 (x-y)^2\Big]$.

Finally and for completeness,
it is instructive to re-express the Hamiltonian in the laboratory frame (\ref{HAM_MIX_CAR})
using the shifts of the coordinates $x_0$ and $y_0$,
\beqn\label{LAB_SHIFT_HAM}
& & \hat H(x_1,\ldots,x_{N_1},y_1,\ldots,y_{N_2}) =
 \sum_{j=1}^{N_1} \left[ -\frac{1}{2m_1} \frac{\partial^2}{\partial x_j^2} +
 \frac{1}{2} m_1\omega^2 \left(x_j-x_0\right)^2 \right] + \nonumber \\ 
& & + \sum_{j=1}^{N_2} \left[ -\frac{1}{2m_2} \frac{\partial^2}{\partial y_j^2} +
\frac{1}{2} m_2\omega^2 \left(y_j-y_0\right)^2 \right] +
\lambda_1 \sum_{1 \le j < k}^{N_1} \left[\left(x_j-x_0\right)-\left(x_k-x_0\right)\right]^2 + \nonumber \\
& & + \lambda_2 \sum_{1 \le j < k}^{N_2} \left[\left(y_j-y_0\right)-\left(y_k-y_0\right)\right]^2 +  
\lambda_{12} \sum_{j=1}^{N_1} \sum_{k=1}^{N_2} \left[\left(x_j-x_0\right)-\left(y_k-y_0\right)\right]^2 + \nonumber \\
& & + \frac{1}{2}m_1N_1 \omega^2 x_0^2 + \frac{1}{2}m_2N_2 \omega^2 (y_0-L)^2 + \lambda_{12}N_1N_2\left(x_0-y_0\right)^2.
\eeqn
Equation (\ref{LAB_SHIFT_HAM}) admits an appealing physical interpretation of demixing and its energetics,
within our model, as discussed in the main text.

\end{document}